\documentclass[
12pt,
preprint,preprintnumbers,nofootinbib,
groupedaddress,superscriptaddress,amsmath,amssymb]{revtex4}
\usepackage{graphicx}
\usepackage{dcolumn}
\usepackage{bm}
\usepackage{amssymb}
\usepackage{amsmath}
\usepackage{epsfig}    
\usepackage{color}
\usepackage{hhline}

\def\be{\begin{equation}}
\def\ee{\end{equation}}
\newcommand{\bea}{\begin{eqnarray}}
\newcommand{\eea}{\end{eqnarray}}
\newcommand{\nn}{\nonumber}

\numberwithin{equation}{section}

\begin{document}
{\begin{flushright}{ APCTP Pre2019 - 017}\end{flushright}}

\title{Modular $S_3$ symmetric radiative seesaw model}

\author{Hiroshi Okada}
\email{hiroshi.okada@apctp.org}
\affiliation{Asia Pacific Center for Theoretical Physics, Pohang 37673, Republic of Korea}
\affiliation{Department of Physics, Pohang University of Science and Technology, Pohang 37673, Republic of Korea}

\author{Yuta Orikasa}
\email{Yuta.Orikasa@utef.cvut.cz}
\affiliation{Institute of Experimental and Applied Physics, 
Czech Technical University in Prague, 
Husova 240/5, 110 00 Prague 1, Czech Republic}

\date{\today}

\begin{abstract}
We propose a one-loop induced radiative seesaw model applying a modular $S_3$ flavor symmetry, which is known as the minimal non-Abelian discrete group. In this scenario, dark matter (DM) candidate is correlated with neutrinos and lepton flavor violations (LFVs). 
 We show several predictions of mixings and phases satisfying LFVs, observed relic density, and neutrino oscillation data.  
\end{abstract}
\maketitle
\newpage

\section{Introduction}
Radiative seesaw models are one of the attractive scenarios to describe {tiny neutrino masses} and dark matter (DM) candidate at the same time~\cite{Ma:2006km}. Subsequently, several phenomenologies such as lepton flavor violations (LFVs), muon anomalous magnetic moment, and collider physics can be taken in account, depending on models.
In addition, modular flavor symmetries have been recently proposed~\cite{Feruglio:2017spp, deAdelhartToorop:2011re}
to provide more predictions to the quark and lepton sector due to Yukawa couplings with a representation of a group.  
Their typical groups are found in basis of  the $A_4$ modular group \cite{Feruglio:2017spp, Criado:2018thu, Kobayashi:2018scp, Okada:2018yrn, Nomura:2019jxj, Okada:2019uoy, deAnda:2018ecu, Novichkov:2018yse, Nomura:2019yft, Ding:2019xvi, Okada:2019mjf, Nomura:2019lnr, Kobayashi:2019xvz, Asaka:2019vev, Gui-JunDing:2019wap, Zhang:2019ngf}, 
$S_3$ \cite{Kobayashi:2018vbk, Kobayashi:2018wkl, Kobayashi:2019rzp}, 
$S_4$ \cite{Penedo:2018nmg, Novichkov:2018ovf, Kobayashi:2019mna, King:2019vhv, Okada:2019lzv, Criado:2019tzk, Kobayashi:2019xvz, Gui-JunDing:2019wap, Wang:2019ovr}, 
$A_5$ \cite{Novichkov:2018nkm, Ding:2019xna, Criado:2019tzk}, larger groups~\cite{Baur:2019kwi}, multiple modular symmetries~\cite{deMedeirosVarzielas:2019cyj}, and double covering of $A_4$~\cite{Liu:2019khw} in which  masses, mixings, and CP phases for quark and lepton are predicted.~\footnote{Several reviews are helpful to understand whole the ideas~\cite{Altarelli:2010gt, Ishimori:2010au, Ishimori:2012zz, Hernandez:2012ra, King:2013eh, King:2014nza, King:2017guk, Petcov:2017ggy, Xing:2019vks} for traditional applications and \cite{Baur:2019iai, Kobayashi:2019uyt} for modular symmetries.}
Furthermore, thanks to the modular weight that is another degree of freedom originated from modular symmetry,
 this modular weight can be identified as a symmetry to stabilize DM candidate if DM is included in a model.
 Thus, radiative seesaw models with modular flavor symmetries are well motivated in view of neutrino predictions and DM origin.

In this paper, we apply a $S_3$ modular symmetry to the lepton sector in a framework of Ma model~\cite{Ma:2006km}, where $S_3$ is known as the minimal symmetry in non-Abelian discrete flavor symmetry. Here, we introduce two right-handed neutrinos that correspond to two singlets under $S_3$ and an isospin doublet inert boson in standard model (SM), both of which have nonzero charge of modular weight.
In order to get a radiative seesaw model, we introduce additional $Z_2$ symmetry since the modular invariance is not sufficient to retain the radiative seesaw model. Therefore, $Z_2$ plays an role in assuring stability of DM.
However, we realize a neutrino predictive model under one of the active neutrino masses is vanishing due to the two right-handed Majorana fermions, where the two kinds of fields originate from the fact that there are only two singlets under $S_3$.~\footnote{If we assign the right-handed Majorana fields as doublet under $S_3$, we cannot reproduce the observed neutrino oscillation data because of few free parameters.}
This is the first achievement in several series of modular flavor symmetry projects.

In our analysis, we show several predictions to the lepton sector, satisfying constraints of LFVs as well as neutrino oscillation data. Also, bosonic DM is favored compared to the fermionic one, since the interacting coupling between DM and the SM particles are too tiny to explain the observed relic density.~\footnote{Another stabilization mechanism of DM candidate has been discussed in non-Abelian discrete symmetries in Refs.~\cite{Hirsch:2010ru, Lamprea:2016egz, delaVega:2018cnx}.}

This paper is organized as follows.
In Sec.~\ref{sec:realization},   we give our model set up under modular $S_3$ symmetry.
Then, we discuss right-handed neutrino mass spectrum, lepton flavor violation (LFV), 
relic density of DM and generation of the active neutrino mass at one loop level.
Finally we conclude and discuss in Sec.~\ref{sec:conclusion}.

\section{Model} 
\label{sec:realization}
The modular group $\bar\Gamma$ is the group of linear fractional transformation
$\gamma$ acting on the modulus  $\tau$, 
belonging to the upper-half complex plane as:
\begin{equation}\label{eq:tau-SL2Z}
\tau \longrightarrow \gamma\tau= \frac{a\tau + b}{c \tau + d}\ ,~~
{\rm where}~~ a,b,c,d \in \mathbb{Z}~~ {\rm and }~~ ad-bc=1, 
~~ {\rm Im} [\tau]>0 ~ ,
\end{equation}
 which is isomorphic to  $PSL(2,\mathbb{Z})=SL(2,\mathbb{Z})/\{I,-I\}$ transformation.
This modular transformation is generated by $S$ and $T$, 
\begin{eqnarray}
S:\tau \longrightarrow -\frac{1}{\tau}\ , \qquad\qquad
T:\tau \longrightarrow \tau + 1\ ,
\end{eqnarray}
which satisfy the following algebraic relations, 
\begin{equation}
S^2 =\mathbb{I}\ , \qquad (ST)^3 =\mathbb{I}\ .
\end{equation}

 We introduce the series of groups $\Gamma(N)~ (N=1,2,3,\dots)$ defined by
 \begin{align}
 \begin{aligned}
 \Gamma(N)= \left \{ 
 \begin{pmatrix}
 a & b  \\
 c & d  
 \end{pmatrix} \in SL(2,\mathbb{Z})~ ,
 ~~
 \begin{pmatrix}
  a & b  \\
 c & d  
 \end{pmatrix} =
  \begin{pmatrix}
  1 & 0  \\
  0 & 1  
  \end{pmatrix} ~~({\rm mod} N) \right \}
 \end{aligned} .
 \end{align}
 For $N=2$, we define $\bar\Gamma(2)\equiv \Gamma(2)/\{I,-I\}$.
Since the element $-I$ does not belong to $\Gamma(N)$
  for $N>2$, we have $\bar\Gamma(N)= \Gamma(N)$,
  which are infinite normal subgroup of $\bar \Gamma$, called principal congruence subgroups.
   The quotient groups defined as
   $\Gamma_N\equiv \bar \Gamma/\bar \Gamma(N)$
  are  finite modular groups.
In this finite groups $\Gamma_N$, $T^N=\mathbb{I}$  is imposed.
 The  groups $\Gamma_N$ with $N=2,3,4,5$ are isomorphic to
$S_3$, $A_4$, $S_4$ and $A_5$, respectively \cite{deAdelhartToorop:2011re}.

Modular forms of  level $N$ are 
holomorphic functions $f(\tau)$  transforming under the action of $\Gamma(N)$ as:
\begin{equation}
f(\gamma\tau)= (c\tau+d)^k f(\tau)~, ~~ \gamma \in \Gamma(N)~ ,
\end{equation}
where $k$ is the so-called as the  modular weight.

We discuss the modular symmetric theory without supersymmetry. 
In this paper, we fix the $S_3$ ($N=2$) modular group. 
Under the modular transformation of Eq.(\ref{eq:tau-SL2Z}), fields $\phi^{(I)}$ 
transform as 
\begin{equation}
\phi^{(I)} \to (c\tau+d)^{-k_I}\rho^{(I)}(\gamma)\phi^{(I)},
\end{equation}
where  $-k_I$ is the modular weight and $\rho^{(I)}(\gamma)$ denotes an unitary representation matrix of $\gamma\in\Gamma(2)$.

The kinetic terms of their scalar fields are written by 
\begin{equation}
\sum_I\frac{|\partial_\mu\phi^{(I)}|^2}{(-i\tau+i\bar{\tau})^{k_I}} ~,
\label{kinetic}
\end{equation}
which is invariant under the modular transformation.
Also, the Lagrangian should be invariant under the modular symmetry.


\begin{center} 
\begin{table}[tb]
\begin{tabular}{|c||c|c|c|c|c|c||c|c||}\hline\hline  
&\multicolumn{6}{c||}{ Fermions} & \multicolumn{2}{c||}{Bosons} \\\hline
  & ~$\bar L_{L_e}$~& ~$\bar L_{L_2}\equiv(\bar L_{L_\mu},\bar L_{L_\tau})^T$~ & ~$e_{R_e}$~& ~$e_{R_2}\equiv(e_{R_{\mu}},e_{R_{\tau}})^T$~& ~$N_{R_1}$~ 
  & ~$N_{R_2}$~ & ~$H$~  & ~$\eta^*$~
  \\\hline 
 $SU(2)_L$ & $\bm{2}$  & $\bm{2}$  & $\bm{1}$ & $\bm{1}$   & $\bm{1}$  & $\bm{1}$ & $\bm{2}$ & $\bm{2}$    \\\hline 
$U(1)_Y$ & $\frac12$ & $\frac12$ & $-1$  & $-1$& $0$ & $0$  & $\frac12$  & -$\frac12$      \\\hline
 $S_3$ & $1$ & $2$ & $1$ & $2$ & $1$ & $1'$ & $1$ & $1$     \\\hline
 $-k$ & $-2$ & $-2$ & $-2$ & $0$ & $-2$ & $-2$ & $0$ & $-2$   \\\hline
 $Z_2$ & $+$ & $+$ & $+$ & $+$ & $-$ & $-$ & $+$ & $-$   \\\hline
\end{tabular}
\caption{Field contents of fermions and bosons
and their charge assignments under $SU(2)_L\times U(1)_Y\times S_{3}\times Z_2$ in the lepton and boson sector, 
where $-k$ is the number of modular weight 
and the quark sector is the same as the SM.}
\label{tab:fields}
\end{table}
\end{center}

\begin{center} 
\begin{table}[tb]
\begin{tabular}{|c||c|c|c|c|c|c||}\hline\hline  
 &\multicolumn{6}{c||}{Couplings}  \\\hline
  & ~$Y^{(4)}_{\bf1}$~& ~$Y^{(6)}_{\bf1}$~ & ~$Y^{(6)}_{\bf1'}$~ & ~$Y^{(2)}_{\bf2}$~& ~$Y^{(4)}_{\bf2}$~& ~$Y^{(6)}_{\bf2}$~  \\\hline 
 $S_3$ & ${\bf1}$ & ${\bf1}$& ${\bf1}$ & ${\bf2}$ & ${\bf2}$ & ${\bf2}$    \\\hline
 $-k$ & $4$ & $6$ & $6$ & $2$ & $4$& $6$   \\\hline
\end{tabular}
\caption{Modular weight assignments for Yukawa interaction.
}
\label{tab:couplings}
\end{table}
\end{center}

Here, we describe our scenario based on the Ma model, where 
field contents are exactly the same as the Ma model~\cite{Ma:2006km}. 
The {$S_3$ representation} and modular weight
are given by Table~\ref{tab:fields}, while the ones of Yukawa couplings are given by Table~\ref{tab:couplings}.
Under these symmetries, one writes renormalizable Lagrangian as follows:
\begin{align}
-{\cal L}_{Lepton} &=
\alpha_\ell (Y^{(2)}_{\bf 2}\otimes\bar L_{L_2}\otimes e_{R_2})_{\bf1}H
+\beta_\ell (Y^{(4)}_{\bf 2}\otimes\bar L_{L_2}\otimes e_{R_e})_{\bf1}H\nn\\
&+\gamma_\ell (Y^{(2)}_{\bf 2}\otimes\bar L_{L_e}\otimes e_{R_2})_{\bf1}H
+\sigma_\ell (Y^{(4)}_{\bf 1}\otimes\bar L_{L_e}\otimes e_{R_e})_{\bf1}H\nn\\
&+\alpha_\nu (Y^{(6)}_{\bf 2}\otimes\bar L_{L_2}\otimes N_{R_2})_{\bf1}\tilde\eta
+\beta_\nu (Y^{(6)}_{\bf 1'}\otimes\bar L_{L_e}\otimes N_{R_2})_{\bf1}\tilde\eta
\nn\\
&+\rho_\nu (Y^{(6)}_{\bf 1}\otimes\bar L_{L_e}\otimes N_{R_1})_{\bf1}\tilde\eta
+\sigma_\nu (Y^{(6)}_{\bf 2}\otimes\bar L_{L_2}\otimes N_{R_1})_{\bf1}\tilde\eta\nn\\
&+ M_0 (Y^{(4)}_{\bf 1} \otimes\bar N^C_{R_1}\otimes N_{R_1})_{\bf1}
+ M_1 (Y^{(4)}_{\bf 1} \otimes\bar N^C_{R_2}\otimes N_{R_2})_{\bf1}
+ {\rm h.c.}, \label{eq:lag-lep}
\end{align}
where $\tilde\eta\equiv i\sigma_2 \eta^*$, $\sigma_2$ being second Pauli matrix.

The  modular forms with the lowest weight 2; $Y^{(2)}_{\bf2}\equiv (y_1,y_2)$, transforming
as a doublet of $S_3$ is written in terms of Dedekind eta-function  $\eta(\tau)$ and its derivative \cite{Novichkov:2019sqv}:
\begin{eqnarray} 
\label{eq:Y-S3}
y_1(\tau) &=& \frac{i}{4\pi}\left( \frac{\eta'(\tau/2)}{\eta(\tau/2)}  +\frac{\eta'((\tau +1)/2)}{\eta((\tau+1)/2)}  
- \frac{8\eta'(2\tau)}{\eta(2\tau)}  \right), \nonumber \\
y_2(\tau) &=& \frac{\sqrt3 i}{4\pi}\left( \frac{\eta'(\tau/2)}{\eta(\tau/2)}  -\frac{\eta'((\tau +1)/2)}{\eta((\tau+1)/2)}  
 \right) \label{Yi}.
\end{eqnarray}
%
Then, any couplings of higher weight are constructed by multiplication rules of $S_3$,
and one finds the following couplings:
\begin{align}
&Y^{(4)}_{\bf1}=y^2_1+y^2_2,\quad
Y^{(6)}_{\bf1}=3y^2_1y_2-y^3_2,\quad
Y^{(6)}_{\bf1'}=y_1^3-3y_1y_2^2,\nn\\
&Y^{(4)}_{\bf2}=
\left[\begin{array}{c}
2y_1y_2 \\ 
y^2_1-y_2^2  \\ 
\end{array}\right],\quad
Y^{(6)}_{\bf2}=
\left[\begin{array}{c}
y^3_1+y_1y_2^2 \\ 
y^3_2+y_1^2 y_2 \\ 
\end{array}\right]. 
\end{align}

Higgs potential is given by
\begin{align}
{\cal V} &= -\mu_H^2 |H|^2 +\mu^2_\eta |Y^{(4)}_{\bf1}||\eta|^2\\
&+ \frac14 \lambda_H|H|^4+ \frac14\lambda_\eta |Y^{(8)}_{\bf1}| |\eta|^4
+\lambda_{H\eta} |Y^{(4)}_{\bf1}||H|^2|\eta|^2+\lambda_{H\eta}' |Y^{(4)}_{\bf1}| |H^\dag\eta|^2
+\frac14\lambda_{H\eta}'' [Y^{(4)}_{\bf1}(H^\dag\eta)^2+ {\rm h.c.}]\nn,
 \label{eq:pot}
\end{align}
which can be the same as the original potential of Ma model without loss of generality, because of additional free parameters.
The point is that one does not have a term $H^\dag\eta$ due to absence of $S_3$ singlet with modular weight $2$ that arises from the feature of modular symmetry.

{The structure of Yukawa couplings are determined by the modular symmetry. Therefore, our model is more predictive 
than the standard Ma model. }
After the electroweak spontaneous symmetry breaking,  the charged-lepton mass matrix is given by
\begin{align}
m_\ell&= \frac {v_H}{\sqrt{2}}
\left[\begin{array}{ccc}
\sigma_\ell Y_{\rm1}^{(4)} & \gamma_\ell y_1 & \gamma_\ell y_2 \\ 
\beta_\ell (2y_1y_2) & \alpha_\ell y_2 &  \alpha_\ell y_1 \\ 
\beta_\ell (y_1^2-y_2^2) & \alpha_\ell y_1 & -\alpha_\ell y_2 \\ 
\end{array}\right], 
\end{align}
where {$\langle H\rangle\equiv [0, v_H/\sqrt2]^T$}.
Then the charged-lepton mass eigenstate can be found by $|D_\ell|^2\equiv V_{e_L} m_\ell m^\dag_\ell V_{e_L}^\dag$.
In our numerical analysis below, one can numerically fix the free parameters $\alpha_\ell,\beta_\ell,\gamma_\ell$ to fit the three charged-lepton masses after giving all the numerical values. Therefore, $\sigma_\ell$ is an input parameter that is free.

 The right-handed neutrino mass matrix is given by
\begin{align}
{\cal M_N} &=
\left[\begin{array}{cc}
M_0 Y^{(4)}_{\bf1} &0 \\ 
0 & M_1 Y^{(4)}_{\bf1}   \\ 
\end{array}\right].\label{eq:mn}
\end{align}
It suggests that right-handed neutrinos are diagonal with two degenerate masses for the second and third fields, and we define
${M_N}_1\equiv M_0 Y^{(4)}_{\bf1}$, ${M_N}_2  \equiv M_1 Y^{(4)}_{\bf1}$.

The Dirac Yukawa matrix is given by
\begin{align}
y_D &=
\left[\begin{array}{cc}
\rho_\nu  Y^{(6)}_{\bf1} & \beta_\nu  Y^{(6)}_{\bf1'}
 \\ 
\sigma_\nu Y^{(6)}_{\bf2,1} 
 &
- \alpha_\nu   Y^{(6)}_{\bf2,2}     
\\ 
\sigma_\nu Y^{(6)}_{\bf2,2}  & \alpha_\nu   Y^{(6)}_{\bf2,1}  
  \\ 
\end{array}\right],\label{eq:mn}
\end{align}
where $Y^{(6)}_{\bf2}\equiv[Y^{(6)}_{\bf2,1},Y^{(6)}_{\bf2,2}]^T$.

{\it Lepton flavor violations} also arises from $y_D$ as~\cite{Baek:2016kud, Lindner:2016bgg}
\begin{align}
&{\rm BR}(\ell_i\to\ell_j\gamma)\approx\frac{48\pi^3\alpha_{em}C_{ij}}{G_F^2 (4\pi)^4}
\left|\sum_{\alpha=1-3}y_{D_{j\alpha}} y^\dag_{D_{\alpha i}} F(M_{\alpha},m_{\eta^\pm})\right|^2,\\
&F(m_a,m_b)\approx\frac{2 m^6_a+3m^4_am^2_b-6m^2_am^4_b+m^6_b+12m^4_am^2_b\ln\left(\frac{m_b}{m_a}\right)}{12(m^2_a-m^2_b)^4},
\end{align}
where $C_{21}=1$, $C_{31}=0.1784$, $C_{32}=0.1736$, $\alpha_{em}(m_Z)=1/128.9$, and $G_F=1.166\times10^{-5}$ GeV$^{-2}$.
The experimental upper bounds are given by~\cite{TheMEG:2016wtm, Aubert:2009ag,Renga:2018fpd}
\begin{align}
{\rm BR}(\mu\to e\gamma)\lesssim 4.2\times10^{-13},\quad 
{\rm BR}(\tau\to e\gamma)\lesssim 3.3\times10^{-8},\quad
{\rm BR}(\tau\to\mu\gamma)\lesssim 4.4\times10^{-8},\label{eq:lfvs-cond}
\end{align}
which will be imposed in our numerical calculation.
 %

{\it Neutrino mass matrix}  is given at one-loop level by
\begin{align}
&m_{\nu_{ij}}\approx \sum_{\alpha=1,2}\frac{y_{D_{i\alpha}} {M_N}_{\alpha} y^T_{D_{\alpha j}}}{(4\pi)^2}
\left(\frac{m_R^2}{m_R^2-{ M^2_N}_{\alpha}}\ln\left[\frac{m_R^2}{{ M^2_N}_{\alpha}}\right]
-
\frac{m_I^2}{m_I^2-{M^2_N}_{\alpha}}\ln\left[\frac{m_I^2}{{ M^2_N}_{\alpha}}\right]
\right),
\end{align}
where {$m_{R(I)}$ is a mass of the real (imaginary) component of $\eta^0$}.
Then the neutrino mass matrix is diagonalized by an unitary matrix $U_{\nu}$ as $U_{\nu}m_\nu U^T_{\nu}=$diag($m_{\nu_1},m_{\nu_2},m_{\nu_3}$)$\equiv D_\nu$, where Tr$[D_{\nu}] \lesssim$ 0.12 eV is given by the recent cosmological data~\cite{Aghanim:2018eyx}.
Then, one finds $U_{PMNS}=V^\dag_{eL} U_\nu$.
Each of mixing is given in terms of the component of $U_{MNS}$ as follows:
\begin{align}
\sin^2\theta_{13}=|(U_{PMNS})_{13}|^2,\quad 
\sin^2\theta_{23}=\frac{|(U_{PMNS})_{23}|^2}{1-|(U_{PMNS})_{13}|^2},\quad 
\sin^2\theta_{12}=\frac{|(U_{PMNS})_{12}|^2}{1-|(U_{PMNS})_{13}|^2}.
\end{align}
We provide the experimentally allowed ranges for neutrino mixings and mass difference squares at 3$\sigma$ range~\cite{Esteban:2018azc} as follows:
\begin{align}
&\Delta m^2_{\rm atm}=[2.431-2.622]\times 10^{-3}\ {\rm eV}^2,\
\Delta m^2_{\rm sol}=[6.79-8.01]\times 10^{-5}\ {\rm eV}^2,\\
&\sin^2\theta_{13}=[0.02044-0.02437],\ 
\sin^2\theta_{23}=[0.428-0.624],\ 
\sin^2\theta_{12}=[0.275-0.350].\nn
\end{align}

Also, the effective mass for the neutrinoless double beta decay is given by
\begin{align}
m_{ee}=|D_{\nu_1} \cos^2\theta_{12} \cos^2\theta_{13}+D_{\nu_2} \sin^2\theta_{12} \cos^2\theta_{13}e^{i\alpha_{21}}
+D_{\nu_3} \sin^2\theta_{13}e^{i(\alpha_{31}-2\delta_{CP})}|,
\end{align}
where its observed value could be measured by KamLAND-Zen in future~\cite{KamLAND-Zen:2016pfg}.

To achieve numerical analysis, we derive several relations of the normalized neutrino mass matrix as follows:
\begin{align}
&\tilde m_{\nu_{ij}}\equiv \frac{m_{\nu_{ij}}}{k_3}\approx 
\frac{1}{(4\pi)^2}\sum_{\alpha=1-3}
y_{D_{i\alpha}} \tilde k_\alpha y^T_{D_{\alpha j}},\quad \tilde k_\alpha \equiv \frac{k_\alpha}{k_3} ,\nn\\
 k_\alpha&\equiv {M_N}_{\alpha}
\left(\frac{m_R^2}{m_R^2-{M^2_N}_{\alpha}}\ln\left[\frac{m_R^2}{{M^2_N}_{\alpha}}\right]
-
\frac{m_I^2}{m_I^2-{M^2_N}_{\alpha}}\ln\left[\frac{m_I^2}{{M^2_N}_{\alpha}}\right]\right)\nn\\
&\approx 
 {M_N}_{\alpha} \Delta m^2 \left(\frac{ { {M^2_N}_{\alpha} - m_R^2 + {M^2_N}_{\alpha} \ln\left(\frac{m_R^2}{ {M^2_N}_{\alpha}}\right)}}{({M^2_N}_{\alpha} - m_R^2)^2}\right),
 \label{eq:norm-nu}
\end{align}
where the last line is the first order approximation of the small mass difference between $m_R^2$ and $m_I^2$;
$m_R^2-m_I^2=\Delta m^2$.~\footnote{Advantage of this approximation is that $\tilde k_\alpha$ does not depend on $\Delta m$.}
Then the normalized neutrino mass eigenvalues are given in terms of neutrino mass eigenvalues; diag$(\tilde m_{\nu_1}^2,\tilde m_{\nu_2}^2,\tilde m_{\nu_3}^2)={\rm diag}(m_{\nu_1}^2,m_{\nu_2}^2,m_{\nu_3}^2)/k_3^2$. 
 It is found that $k_3^2$ is given by
\begin{align}
k_3^2=\frac{\Delta m^2_{\rm atm}}{\tilde m_{\nu_3}^2 - \tilde m_{\nu_1}^2},\label{eq:k}
\end{align}
where normal hierarchy is assumed and $\Delta m^2_{\rm atm}$ is the atmospheric neutrino mass difference square.
Comparing Eq.(\ref{eq:norm-nu}) and Eq.(\ref{eq:k}), we find $\Delta m^2$ is rewritten by the other parameters as follows:
\begin{align}
\Delta m^2 \approx
 k_3 \left(\frac{ {M_N}_{2} \left[{ {M^2_N}_{2} - m_R^2 + {M^2_N}_{2} \ln\left(\frac{m_R^2}{ {M^2_N}_{2}}\right)}\right] }
{({M^2_N}_{2} - m_R^2)^2}\right)^{-1} .
 \label{eq:dm}
\end{align}
 The solar neutrino mass difference square is also found as 
 \begin{align}
\Delta m^2_{\rm sol}=\Delta m^2_{\rm atm} \frac{\tilde m_{\nu_2}^2 - \tilde m_{\nu_1}^2}{\tilde m_{\nu_3}^2 - \tilde m_{\nu_1}^2},\label{eq:k}
\end{align}
In numerical analysis, this value should be within the experimental result, while $\Delta m^2_{\rm atm}$ is expected to be input parameter.

DM is expected to be an imaginary component of inert scalar $\eta$; $\eta_I$. In order to avoid the oblique parameters, we assume to be {$m_{\eta^\pm} \approx m_I$} for simplicity.
In this case, the mass of DM is uniquely fixed by the observed relic density which suggests it is within $534\pm8.5$ GeV~\cite{Hambye:2009pw}, if the Yukawa coupling is not so large. In fact, tiny Yukawa couplings are requested by satisfying
the data. Thus, we just work on the mass of $\eta$ at this narrow range.

\section{Numerical analysis}
Here, we show numerical analysis to satisfy all of the constraints that we discussed above,
where we work on a basis that the neutrino mass ordering is normal hierarchy.
\footnote{We have checked that the inverted hierarchy is not favored in our model.}
The range of absolute value of the five complex dimensionless parameters $\alpha_\nu,\beta_\nu, \rho_\nu, \sigma_\nu, \sigma_\ell$ are taken to be $[0.01-1]$,while the mass parameters $M_0, M_1$ are of the order [50,500] TeV. 
We have only two right handed neutrino, therefore $m_1$=0 eV and $\alpha_{21}$=0 [deg].

%
%
%

\begin{figure}[tb]\begin{center}
\includegraphics[width=80mm]{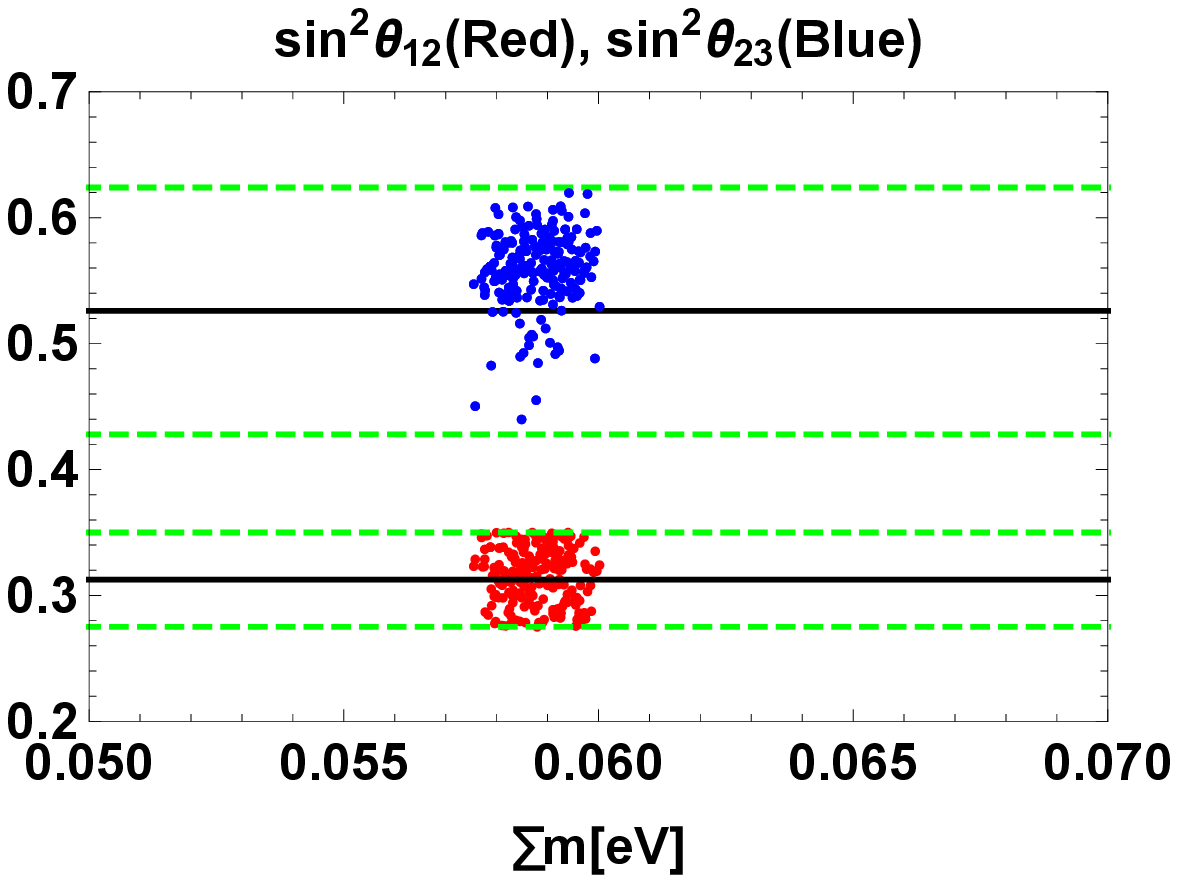}
\includegraphics[width=80mm]{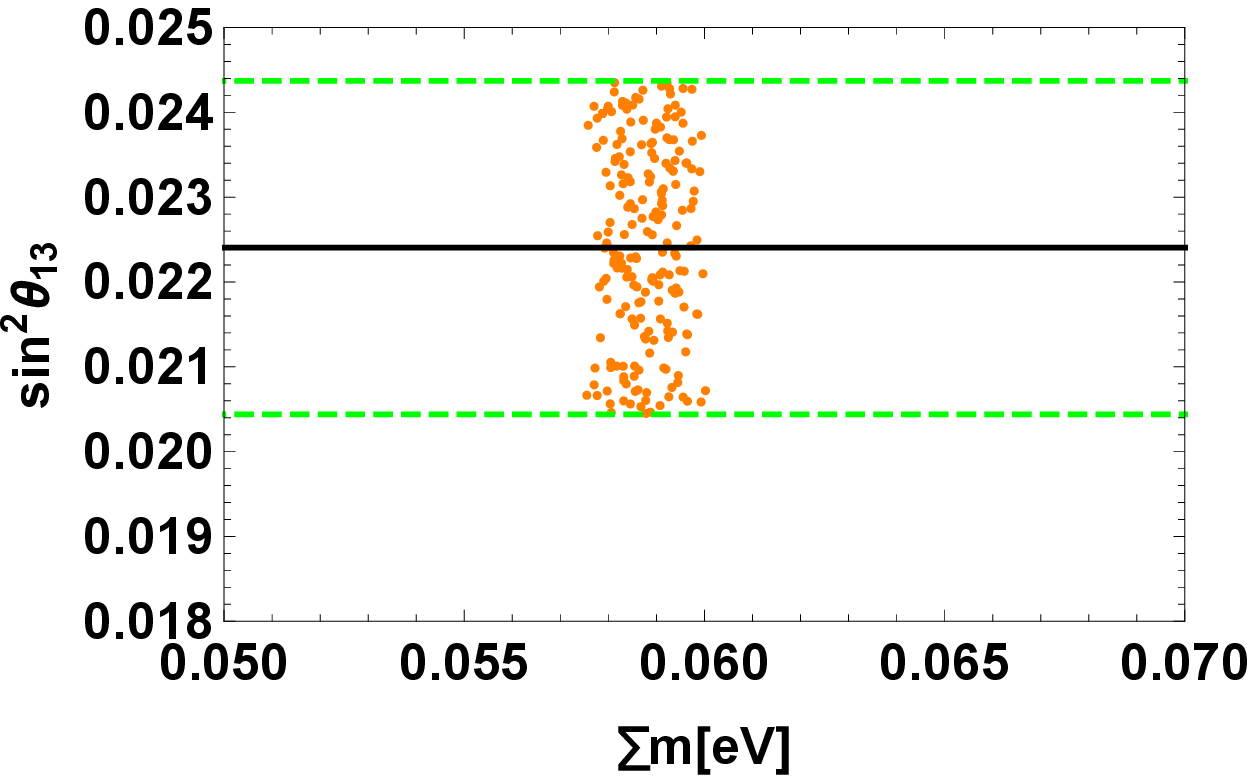}
\caption{The sum of neutrino masses $\sum m(\equiv$ Tr$[ D_\nu]$) versus $\sin^2\theta_{12}$(red color), $\sin^2\theta_{23}$(blue color) in the left figure, and $\sin^2\theta_{13}$(orange color) in the right figure. Here, the horizontal black solid lines are the best fit values, the green dotted lines show 3$\sigma$ range,
and the vertical black line shows upper bound on the cosmological data as shown in the neutrino section.} 
\label{fig:1}\end{center}\end{figure}

Figure~\ref{fig:1} shows the sum of neutrino masses $\sum m(\equiv$ Tr$[D_\nu])$ versus $\sin^2\theta_{12}$(red color), $\sin^2\theta_{23}$(blue color) in the left figure, and $\sin^2\theta_{13}$ in the right figure.
Here, the horizontal black solid lines are the best fit values, the green dotted lines show 3$\sigma$ range,
and the vertical black line shows upper bound on the cosmological data as shown in the neutrino section.
It suggests that all the three mixings run over whole the range of experimental results at 3$\sigma$ interval, 
even though larger value of $\sin^2\theta_{23}$ is somewhat favored. 
While the sum of neutrino masses is restricted to be $\sum m\approx$0.06 eV that always satisfies the upper bound on the cosmological result.

\begin{figure}[tb]\begin{center}
\includegraphics[width=80mm]{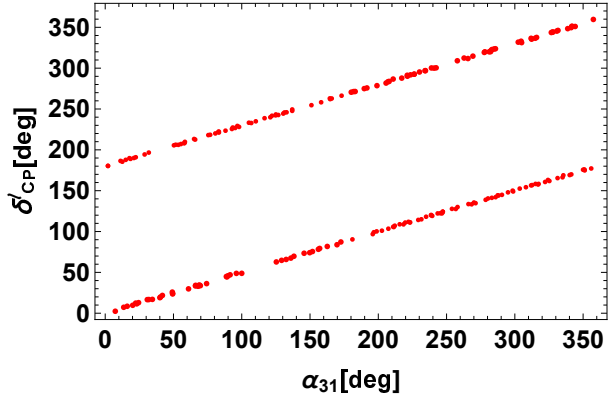}
\caption{Phase of $\delta^\ell_{CP}$ in terms of $\alpha_{31}$.}   
\label{fig:2}\end{center}\end{figure}

Figure~\ref{fig:2} shows phase of $\delta^\ell_{CP}$ in terms of $\alpha_{31}$.
This figure implies that Dirac CP is linearly proportional to $\alpha_{31}$ phase that runs over whole the ranges. 
Once the Dirac CP phase could be fixed to be $\sim270$ [deg] in future experiments, $\alpha_{31}$ is predicted to be $\sim$ 200 [deg].

\begin{figure}[tb]\begin{center}
\includegraphics[width=80mm]{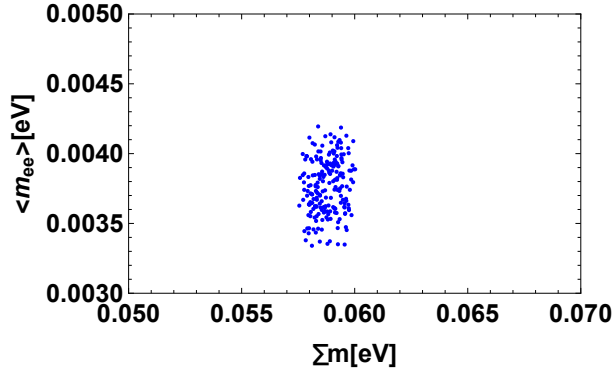}
\caption{The sum of neutrino masses versus the effective mass for the neutrinoless double beta decay.}   
\label{fig:3}\end{center}\end{figure}

Figure~\ref{fig:3} demonstrates the sum of neutrino masses versus {the effective mass for the} neutrinoless double beta decay.
It suggests that 0.0035 eV $\lesssim\langle m_{ee}\rangle\lesssim$ 0.045 eV.
Another remarks are in order:
 \begin{enumerate}
\item
The typical region of modulus $\tau$ is found in narrow space as -0.1\ $\lesssim\ $Re$[\tau]\lesssim$\ 0.1 and  1.2\ $\lesssim\ $Im$[\tau]\lesssim$\ 1.3.
\item 
Typical scale of LFVs are very small in our analyses, therefore following upper bounds are realized:
\[{\rm BR}(\mu\to e\gamma)\lesssim3.0\times10^{-19}, \quad {\rm BR}(\tau\to e\gamma)\lesssim2.5\times10^{-19},\quad {\rm BR}(\tau\to \mu\gamma)\lesssim1.5\times10^{-20}.\]
\item The lightest Majorana mass eigenstate is given by [2$-$9] TeV.
 \end{enumerate}

\section{Conclusion and discussion}
\label{sec:conclusion}
We have constructed a predictive lepton model with modular $S_3$ symmetry in framework of one-loop induced radiative seesaw model. 
The DM stability is naturally assured by $Z_2$ symmetry, and
DM is correlated with neutrinos in a specific manner, where their interactions are determined by the $S_3$ symmetry that is known as the minimal group in non-Abelian discrete flavor symmetries. In our numerical analyses, we have highlighted several remarks as follows: 
 \begin{enumerate}
 \item 
 	The Dirac phase and the Majorana phase are strongly correlated.  
 \item
	The typical region of modulus $\tau$ is found in narrow space as -0.1\ $\lesssim\ $Re$[\tau]\lesssim$\ 0.1 and  1.2\ $\lesssim\ $Im$[\tau]\lesssim$\ 1.3.
 \item 
	Typical scale of LFVs are very small in our analyses, therefore following upper bounds are realized:
\[{\rm BR}(\mu\to e\gamma)\lesssim3.0\times10^{-19}, \quad {\rm BR}(\tau\to e\gamma)\lesssim2.5\times10^{-19},\quad {\rm BR}(\tau\to \mu\gamma)\lesssim1.5\times10^{-20}.\]
 \item 
	The lightest Majorana mass eigenstate is given by [2$-$9] TeV.
 \end{enumerate}

\section*{Acknowledgments}
\vspace{0.5cm}
{\it
This research was supported by an appointment to the JRG Program at the APCTP through the Science and Technology Promotion Fund and Lottery Fund of the Korean Government. This was also supported by the Korean Local Governments - Gyeongsangbuk-do Province and Pohang City (H.O.). H. O. is sincerely grateful for the KIAS member, and log cabin at POSTECH to provide nice space to come up with this project.
Y. O. was supported from European Regional Development Fund-Project Engineering Applications of Microworld
Physics (No.CZ.02.1.01/0.0/0.0/16\_019/0000766)}


\end{document}